\documentclass[%
 superscriptaddress,
 preprint,
 aps,
 prb,
]{revtex4-1}

\usepackage{graphicx}

\begin{document}


\title{Suppression of antiferromagnetic order and hybridization gap 
by electron- and hole-doping in the Kondo semiconductor CeOs$_{2}$Al$_{10}$}

\author{J. Kawabata}
\affiliation{Department of Quantum matter, AdSM, Hiroshima University, Higashi-Hiroshima 739-8530, Japan}
\author{T. Takabatake}
\email[]{takaba@hiroshima-u.ac.jp}
\affiliation{Department of Quantum matter, AdSM, Hiroshima University, Higashi-Hiroshima 739-8530, Japan}
\affiliation{Institute for Advanced Materials Research, Hiroshima University, Higashi-Hiroshima 739-8530, Japan}
\author{K. Umeo}
\email[]{kumeo@sci.hiroshima-u.ac.jp}
\affiliation{Cryogenics and Instrumental Analysis Division, N-BARD, Hiroshima University, Higashi-Hiroshima 739-8526, Japan}
\affiliation{Department of Quantum matter, AdSM, Hiroshima University, Higashi-Hiroshima 739-8530, Japan}
\author{Y. Muro}
\affiliation{Liberal Arts and Sciences, Faculty of Engineering, Toyama Prefectural University, Izumi 939-0398, Japan}

\date{\today}

\begin{abstract}

The Kondo semiconductor CeOs$_{2}$Al$_{10}$ exhibits an antiferromagnetic 
(AFM) order at $T_\mathrm{N}= 28.5$ K, whose temperature is unexpectedly high for 
the small ordered moment of $0.3$ $\mu_\mathrm{B}/$Ce. We have studied the effects 
of electron- and hole-doping on the hybridization gap and AFM order by 
measuring the magnetization $M$, magnetic susceptibility $\chi$, electrical 
resistivity $\rho$, and specific heat $C$ on single crystals of 
Ce(Os$_{1-x}$Ir$_{x}$)$_{2}$Al$_{10}$($x \le 0.15$) and 
Ce(Os$_{1-y}$Re$_{y}$)$_{2}$Al$_{10}$($y \le 0.1$). The results of 
$M (B)$ indicates that the AFM ordered moment $\mu_\mathrm{AF}$ changes the 
direction from the $c$-axis for $x = 0$ to the $a$-axis for $x = 0.03$. With 
increasing $x$ up to 0.15, $T_\mathrm{N}$ gradually decreases although the $4f$ electron 
state becomes localized and the magnitude of $\mu_\mathrm{AF}$ is increased to 
$1$ $\mu_\mathrm{B}/$Ce. With increasing $y$, the $4f$ electron state is more delocalized 
and the AFM order disappears at a small doping level $y = 0.05$. In both 
electron- and hole-doped systems, the suppression of $T_\mathrm{N}$ is well 
correlated with the increase of the Sommerfeld coefficient $\gamma$ in 
$C(T)$. Furthermore, the simultaneous suppression of $T_\mathrm{N}$ and the 
semiconducting gap in $\rho (T)$ at $T > T_\mathrm{N}$ indicates 
that the presence of the hybridization gap is indispensable for the unusual 
AFM order in CeOs$_{2}$Al$_{10}$.

\end{abstract}

\maketitle


\section{\label{sec:level1}INTRODUCTION}
In a few cerium-based compounds, hybridization of $4f$ electrons and conduction 
bands (\textit{c-f} hybridization) gives rise to a hybridization gap in the vicinity of 
the Fermi level.\cite{PS01} Because of the strong electron correlation, the energy 
gap is renormalized to a small value of 10-100 K. So-called Kondo 
semiconductors such as Ce$_{3}$Bi$_{4}$Pt$_{3}$ and CeRhAs display 
semiconducting behavior in the electrical resistivity,\cite{MF02, TT03} while so-called 
Kondo semimetals CeNiSn and CeRhSb show semimetallic behavior.\cite{TT03, TT04} The latter 
behavior reflects the anisotropic gap which closes in a particular direction 
due to the anisotropy of the hybridization.\cite{MK05, HI06, JM07} These compounds belong to the 
valence fluctuating regime and do not order magnetically at low temperatures 
because the strong Kondo interaction between the $4f$ localized moment and 
conduction electron spins screens the localized moments. However, doping of 
$3d$ electrons in CeNiSn by Cu substitution for Ni at a several {\%} level 
induces a long-range antiferromagnetic (AFM) order.\cite{TT08, KN09} The emergence of AFM 
order was attributed to the weakened \textit{c-f} exchange interaction which is a 
consequence of the increase of the Fermi level with respect to the $4f$ level. 
On the other hand, doping of 3$d$ holes in CeNiSn by Co substitution for Ni 
strengthens the \textit{c-f} exchange interaction and thus increases the Kondo 
temperature $T_\mathrm{K}$.\cite{KN09, TT10, DT11} 

A family of compounds CeT$_{2}$Al$_{10}$(T $=$ Fe, Ru, Os) with the 
orthorhombic YbFe$_{2}$Al$_{10}$-type structure display semiconducting 
behavior in the resistivity at high temperatures and thus they were 
classified into Kondo semiconductors.\cite{YM12, AM13, TN14} The Fe compound, where the \textit{c-f} 
hybridization is strongest among the three, belongs to the valence 
fluctuation regime and thus the ground state remains in a paramagnetic 
state.\cite{YM12} However, the isoelectronic compounds with T $=$ Ru and T $=$ Os 
order antiferromagnetically at rather high N\'{e}el temperature $T_\mathrm{N}$ of 27 
K and 28.5 K, respectively.\cite{AM13, TN14, YM15, YM16, HT17, DD18, SK19, DT20, HK21} It evoked a question why the $T_\mathrm{N}$'s of 
the Ce compounds with small magnetic moments $0.3-0.4$ $\mu_\mathrm{B}/$Ce are 
higher than those of Gd counterparts with 7 $\mu_\mathrm{B}/$Gd.\cite{YM16, HT17} The 
ordering temperatures may be scaled by the de Gennes factor as long as the 
magnetic order is caused by the Ruderman -- Kittel -- Kasuya -- Yosida (RKKY) 
interaction.\cite{KN22} According to the de Gennes scaling, $T_\mathrm{N}$ for a Ce compound 
is expected to be 1/100 of that for the Gd counterpart if we neglect the 
crystal field effect and possible difference in the Fermi surface. Optical 
conductivity measurements for CeT$_{2}$Al$_{10}$(T $=$ Ru, Os) have 
revealed the CDW-like instability which develops along the $b$ axis at 
temperatures slightly higher than $T_\mathrm{N}$. It was suggested that this 
electronic instability induces the AFM order.\cite{SK23} Another enigma is why the 
ordered moments $\mu_\mathrm{AF}$ point along the $c$ axis although the magnetic 
susceptibility is largest for $B \parallel a$; $\chi (B \parallel a) \gg \chi 
(B \parallel c) > \chi (B \parallel b)$.\cite{HT17, DD18, SK19, DT20, HK21}

As mentioned above, previous studies of the Kondo semimetal CeNiSn by doping 
$3d$ electrons and holes provided us with important information on the relation 
between the hybridization gap and magnetism. For better understanding of the 
unusual magnetic order in CeOs$_{2}$Al$_{10}$, we have conducted a 
systematic study by substituting Ir and Re for Os, which dope $5d$ electrons 
and $5d$ holes, respectively. A part of the experimental results on the Ir 
substituted samples has been reported in a proceeding of a conference.\cite{JK24}

\section{PREPARATION AND CHARACTERIZATION OF SINGLE CRYSTALS}
Homogeneity ranges in Ce(Os$_{1-x}$Ir$_{x}$)$_{2}$Al$_{10}$ and 
Ce(Os$_{1-y}$Re$_{y}$)$_{2}$Al$_{10}$ were examined by preparing 
polycrystalline samples with the nominal compositions of Ir and Re up to 
50{\%}. The arc melted samples were annealed in an evacuated quartz ampoule 
at $850$ $^{ \circ}$C for 7 days. The samples were characterized by combining 
metallographic examination, powder x-ray diffraction, and wavelength 
dispersive electron-probe microanalysis (EPMA). The Re-substituted samples 
were found to be homogeneous with the composition close to the initial 
one. However, the bottom part of the Ir substituted sample contains an Ir 
rich phase of CeOsIr$_{3}$Al$_{15}$ and the Ir composition in the upper part 
is smaller than the initial one. With the initial composition of $X = 0.5$, 
for example, the real composition $x$ was 0.38. We note that the presence of the 
compound CeIr$_{2}$Al$_{10}$ was not reported but there is a report on 
CeRe$_{2}$Al$_{10}$ with a distinct structure from that of 
CeOs$_{2}$Al$_{10}$.\cite{AS25} X-ray diffraction analysis showed that the 
YbFe$_{2}$Al$_{10}$-type structure is kept in the whole ranges $x \le  0.4$ 
and $y \le 0.5$. The lattice parameters are plotted in Fig. 1. Because the 
change is smaller than 0.3{\%} for $x \le 0.2$ and $y \le 0.1$, we expect that 
the chemical pressure effect on the \textit{c-f} hybridization may be much weaker than 
that of doping of $5d$ electrons and holes in the above composition range.

Single crystalline samples were grown using an Al self-flux method as 
reported previously.\cite{YM15} Alloys of Ce(Os$_{1-X}$Ir$_{X}$)$_{2}$ and 
Ce(Os$_{1-Y}$Re$_{Y}$)$_{2}$ were prepared by arc melting of pure 
elements. The crushed alloy ingots together with an excess amount of Al in 
the composition of $1:2:30$ were loaded into an aluminum crucible, which was 
sealed in a quartz ampoule under an Ar atmosphere of 1/3 atm. The ampoule 
was heated to $1200$ $^{ \circ}$C, kept for 5 hours, and slowly cooled at a 
ratio $2$ $^{ \circ}$C/h to $720$ $^{ \circ}$C, at which temperature the molten 
Al flux was separated by centrifuging. Several crystals of approximately 
$2 \times 2 \times 3$ mm$^{3}$ were obtained. The atomic composition was 
determined by EPMA. The real compositions of Ir $(x)$ were found to be 0, 0.03, 0.04, 
0.08, and 0.15 for the initial ones $X=0, 0.02, 0.03, 0.10$, and 0.20, 
respectively, while the compositions of Re $(y)$ were same as the initial ones 
$Y = 0.01, 0.02, 0.03, 0.05$, and 0.1. The difference between the values of 
$x$ and $X$ is attributed to the segregation of a small amount of impurity phase 
of (Os, Ir)Al$_{4}$. For the measurements of physical properties, we 
carefully avoided the parts containing impurity phases. After the 
single-crystal nature was verified by the Laue back diffraction method, the 
crystals were cut in an appropriate shape for the measurements. 

\section{MAGNETIC, TRANSPORT, AND THERMAL PROPERTIES}
\subsection{Magnetic susceptibility and magnetization}
Using single crystalline samples mentioned above, we have studied the 
magnetic, transport, and thermal properties. The measurement of magnetic 
susceptibility $\chi (T)$ was performed in an external field $B = 1$ T from 
1.8 K to 300 K with a Quantum Design Magnetic Property Measurements System 
(MPMS). The isothermal magnetization $M(B)$ was measured up to $B = 14$ T by the 
dc extraction method using a Quantum Design Physical Property Measurement 
System (PPMS). An ac four-probe method was used for the electrical 
resistivity $\rho (T)$ measurements from 2.6 K to 300 K. The specific heat 
$C(T)$ was measured from 2 K to 300 K by the relaxation method in PPMS. 

Figures 2(a) and 2(b) display, respectively, the variations of $\chi (T)$ for 
Ce(Os$_{1-x}$Ir$_{x}$)$_{2}$Al$_{10}$($x \le 0.15$) and 
Ce(Os$_{1-y}$Re$_{y}$)$_{2}$Al$_{10}$($y \le 0.1$) along the three 
principal axes. For the undoped sample, $\chi_{a}(T)$ in $B \parallel a$ passes through 
a maximum at around 45 K and drops at $T_\mathrm{N}= 28.5$ K. On going from $x 
= 0$ to 0.15, the broad maximum changes to a sharp peak, whose temperature 
decreases to 7 K. Thereby, the value at the maximum increases by 5 times, 
leading to the enhancement of anisotropy, $\chi_{a} \gg \chi_{c} > \chi_{b}$. It is worth noting 
that the data set of $\chi_{i}(T) (i = a$, $b$, $c)$ for $x = 0.15$ at $T > 30$ K 
are in agreement with the calculation taking account of 
the crystal field effect on the localized $4f$ state of Ce$^{3+}$ ion.\cite{KY26, FS27} The 
solid lines for $x = 0.15$ in Fig. 2(a) represent the calculations.\cite{FS27} 
The observed localization of the $4f$ electron state is a consequence of the increase 
in the Fermi level by electron doping, as found in Cu doping in CeNiSn.\cite{TT08, TT10}  The 
inverse of $\chi_{a}(T)$ is plotted versus $T$ in the inset. The Curie-Weiss 
fit to the data between 200 and 300 K gives the paramagnetic Curie 
temperature $\theta_\mathrm{p}$ which value changes from $-20$ K for $x =$ 0 to 
$+26$ K for $x = 0.15$. For $B \parallel b$ and $B \parallel c$, the decrease 
in $\chi (T)$ at $T < T_\mathrm{N}$ disappears with increasing $x$. For $x \ge 0.08$, the sharp peak in 
$\chi_{a}(T)$ at $T_\mathrm{N}$ and the absence of the drop in $\chi_{b}(T)$ and 
$\chi_{c}(T)$ at $T < T_\mathrm{N}$ suggest the AFM ordered moments $\mu 
_\mathrm{AF}$ oriented parallel to the easy $a$ axis. This reorientation of $\mu 
_\mathrm{AF}$ from $\parallel c$-axis to $ \parallel a$-axis is confirmed by the isothermal 
magnetization and neutron diffraction measurements, as will be presented 
below.

Compared with the tendency of localization of $4f$ electrons in 
Ce(Os$_{1-x}$Ir$_{x}$)$_{2}$Al$_{10}$, an opposite trend  
was observed for Ce(Os$_{1-y}$Re$_{y}$)$_{2}$Al$_{10}$. As shown in Fig. 
2(b), the maximum in $\chi_{a}(T)$ at around 45 K is suppressed with 
increasing $y$ and the anomaly at $T_\mathrm{N}$ disappears at $y = 0.05$. The anomalies 
in $\chi_{b}(T)$ and $\chi_{c}(T)$ disappear at $y = 0.01$ and $y = 0.03$, 
respectively. The continuous increase in $\chi_{a}(T)$ on cooling to 2 K 
may be the effect of disorder-induced magnetic moments, which needs to be 
studied by microscopic experiments. The inset of Fig. 2(b) displays the 
inverse of $\chi_{a}(T)$ against $T$. The increase of $|\theta 
_\mathrm{p}|$ from 20 K for $y = 0$ to 46 K for $y = 0.1$ suggests the 
increase of $T_\mathrm{K}$ because the value of $T_\mathrm{K}$ for the overall 
crystal-field levels is in proportion to $|\theta 
_\mathrm{p}|$.\cite{NB28}

In the undoped sample, $\mu_\mathrm{AF}$ is oriented along the $c$-axis.\cite{DT20, HK21} When 
external magnetic field was applied along the $c$-axis at 0.3 K, a spin flop 
transition was observed at 6 T.\cite{YM15} This transition in $M(B \parallel c)$ is found at 
$B_{sf} = 6.1$ T by the present measurement at 2 K as shown at the bottom 
of Fig. 3(a). For $x = 0.03$, however, there is no transition in 
$M(B \parallel c)$ but a weak upturn appears in $M(B \parallel a)$ at around 3 T. 
For $x = 0.08$, a metamagnetic transition in $M(B \parallel a)$ is clearly seen at 3 T. Passing through a 
bend at $B_{s} = 10.4$ T, $M(B \parallel a)$ is saturated to a large value of 0.7 $\mu 
_\mathrm{B}/$Ce. We interpret the metamagnetic transition along the easy $a$-axis as 
a spin-flop transition from $\mu_\mathrm{AF} \parallel a$ to $\mu_\mathrm{AF} \bot a $ because 
the linear extrapolation of the $M(B \parallel a)$ data between 8 T and 4 T goes to the 
origin. For $x = 0.15$, the magnitude of $M(B \parallel a)$ increases further and that of 
$B_{s}$ decreases, but the spin-flop transition field $B_{sf}$ does not 
change. These observations suggest that the ground state for $x \ge 0.08$ is 
the AFM state with $\mu_\mathrm{AF}$ $(\sim 1 \mu_\mathrm{B}/$Ce) 
pointing along the $a$-axis. Figure 3(b) shows the results of $M(B \parallel a)$ and 
$M(B \parallel c)$ for the Re substituted samples. With increasing $y$, the metamagnetic 
behavior in $M(B \parallel c)$ is barely observed for $y = 0.03$. It is not seen for $y = 0.05$, in 
agreement with the absence of anomaly in $\chi_{a}(T)$ in Fig. 2(b). The 
absence of anomaly in $M(B \parallel a)$ suggests that the AFM order with 
$\mu_\mathrm{AF} \parallel c$ fades away without showing reorientation. 

Figure 4 shows the variations of $B_{sf}$ and $B_{s}$ as a function of $x$ and 
$y$. As mentioned above, the AFM structure changes from $\mu_\mathrm{AF} \parallel c$ to $\mu 
_\mathrm{AF} \parallel a$ somewhere between $x = 0$ and $x = 0.03$. Thereby, the change in 
$T_\mathrm{N}$ is only 10{\%} from 28.5 K to 26.5 K, suggesting that the intersite 
AFM interaction between Ce moments may not depend on the direction of 
$\mu_\mathrm{AF}$ with respect to the crystal axis. Recently, a neutron 
scattering experiment has confirmed the reorientation of ordered moments 
from $\mu_\mathrm{AF} \parallel c = 0.3$ $\mu_\mathrm{B}/$Ce for $x = 0$ to $\mu_\mathrm{AF} 
\parallel a = 0.9$ $\mu_\mathrm{B}/$Ce for $x = 0.08$.\cite{DD29} Similar reorientation of AFM 
ordered moments has been suggested to occur in CeRu$_{2}$Al$_{10}$ when Rh 
is partially substituted for Ru at 5{\%}.\cite{AK30} Since both substitutions of Ir 
for Os and Rh for Ru dope $d$ electrons in the mother compounds, the spin 
reorientation should be induced by the doping of $d$ electrons into the gapped 
state. The magnitude of $B_{s}$ slightly decreases but that of $B_{sf}$ is 
constant in the range $0.03 \le x \le 0.15$. In contrast, $B_{sf}$ is 
strongly suppressed by Re substitution. This result is consistent with the 
neutron scattering study on the sample with $y = 0.03$ which has revealed the 
AFM arrangement of $\mu_\mathrm{AF} \parallel a$ with a reduced size of 0.18 $\mu_\mathrm{B}/$Ce.\cite{DD31} This 
strong suppression of $\mu_\mathrm{AF}$ by Re substitution indicates that even 
low level doping of $5d$ holes in CeOs$_{2}$Al$_{10}$ enhances the Kondo effect 
and prevents the system from AFM ordering.

\subsection{Electrical resistivity}
Effects of the doping on the hybridization gap above $T_\mathrm{N}$ and the AFM gap 
below $T_\mathrm{N}$ may manifest themselves in the temperature dependence of 
electrical resistivity $\rho (T)$. Figures 5(a) and 5(b) show the results of 
$\rho (T)$ along the three principal axes for CeOs$_{2}$Al$_{10}$ 
and substituted samples with Ir and Re, respectively. The vertical lines 
denote $T_\mathrm{N}$'s determined by the specific heat measurement as will be 
described below. The $\rho (T)$ curves for $x = 0$ strongly increase on cooling 
as manifested by the ratio $\rho ($2.6 K$)/ \rho ($300 K$) = 7$ which is 
higher than that of 1.6 for the previously reported data.\cite{YM15} This fact 
indicates higher quality of the present sample. The --log$T$ dependence from 
300 K to 100 K is followed by a thermal activation-type behavior in the 
range from 60 K to 30 K as shown in the inset. By fitting the data with the formula $\rho = 
\rho_{0}$exp($ \mit \Delta$$/2k_\mathrm{B}T)$, the values of 
$\mit \Delta_{a}/k_\mathrm{B}$, $\mit \Delta_{b}/k_\mathrm{B}$, and $\mit \Delta 
_{c}/k_\mathrm{B}$ are estimated to be 56 K, 83 K, and 65 K, respectively, 
whose values are larger by 1.8 times than those reported previously.\cite{YM15} The 
slope of $\rho (T)$ increases abruptly below $T_\mathrm{N}= 28.5$ K, which can 
be attributed to the formation of a superzone gap on the Fermi surface. Such 
a superzone gap is formed by folding of the Brillouin zone associated with 
the AFM order.\cite{HM32} The semiconducting behavior at $T < 14$ K suggests 
opening of another gap within the hybridization gap. Successive openings of 
gaps at the Fermi level have been directly observed by a photoemission 
spectroscopic study.\cite{TI33} At a small level of $x = 0.04$, the semiconducting 
increase at $T < 14$ K changes to a gradual decrease on cooling, 
although both the bend at $T_\mathrm{N}$ and the activation-type behavior above 
$T_\mathrm{N}$ are still observed. The magnitudes of $\mit \Delta_{a}$ , $\mit \Delta 
_{b}$, and $\mit \Delta_{c}$ for $x = 0.04$ are approximately 80{\%} of those 
for $x = 0$. For $x = 0.08$, only $\rho_{a}(T)$ displays the bend at 
$T_\mathrm{N}$, which is consistent with the AFM arrangement of $\mu_\mathrm{AF} \parallel a$ 
along the propagation vector (1, 0, 0).\cite{DD29} In addition, a hump manifests 
itself in $\rho_{c}(T)$ at around 200 K. The --log$T$ behavior above the hump 
can be attributed to the Kondo scattering in the crystal-field excited 
state.\cite{BC34} For $x = 0.15$, the hump at around 200 K becomes more evident in 
$\rho_{c}(T)$ and $\rho_{a}(T)$, but the anomaly at $T_\mathrm{N}$ becomes 
unclear. 

Let us turn our attention to the results of $\rho (T)$ for the Re substituted 
samples in Fig. 5(b). Even at a small level $y = 0.01$, the semiconducting 
behavior in $\rho (T)$ at $T < 10$ K disappears although the thermal 
activation behavior above $T_\mathrm{N}$ still exists. On going from $y = 0.01$ to 
$y = 0.05$, the metallic behavior at low temperatures becomes more evident. 
The anomaly at $T_\mathrm{N}$ is observed in $\rho_{b}(T)$ for $y = 0.02$ but is 
hardly observed for $y = 0.03$. For $y = 0.05$, the maximum shifts to high 
temperature and $\rho (T)$ at $T < 15$ K is in proportion to $T^{2}$. 
For $y = 0.1$, the temperature at the maximum further increases, 
suggesting the enhancement of $T_\mathrm{K}$. It is noteworthy that  the thermal 
activation behavior above $T_\mathrm{N}$ disappears simultaneously with the metalization below $T_\mathrm{N}$. 
This transformation by $5d$ hole doping occurs at a lower doping level than by $5d$ electron doping.

\subsection{Specific heat}
The data of specific heat divided by temperature $C/T$ are plotted against $T$ in 
Figs. 6(a) and 6(b). The midpoint of the jump in $C/T$ was taken as $T_\mathrm{N}$. For 
the undoped sample, $C/T$ jumps at $T_\mathrm{N}$ and the extrapolation of the plot of 
$C/T$ vs $T^{2}$ to $T = 0$ gives Sommerfeld coefficient $\gamma $ of 0.007 
J/K$^{2}$mol. When $x$ is increased to 0.04 and 0.08, $T_\mathrm{N}$ decreases and the 
jump becomes smaller. For $x = 0.15$, $C/T$ gradually increases on cooling from 
17 K and exhibits a jump at 7 K, which temperature agrees with that of the 
sharp peak in $\chi_{a}(T)$ in Fig. 2(a). As is shown in Fig. 6(b), with 
increasing $y$, the jump at $T_\mathrm{N}$ gradually decreases and disappears at $y = 
0.05$. The $\gamma $ value increases to 0.1 J/K$^{2}$mol at $y = 0.1$. 

The values of $T_\mathrm{N}$ and $\gamma$ are plotted as a function of $x$ and $y$ in 
Fig. 7(a). As $5d$ electrons or holes are doped, the value of $T_\mathrm{N}$ is 
suppressed and the $\gamma$ value is increased. The opposite change in 
$T_\mathrm{N}$ and $\gamma$ indicates that the development of the density of the 
states at the Fermi level destroys the AFM order. This fact seems to be 
inconsistent with the AFM order caused by the RKKY mechanism, in which the 
magnetic interaction between neighboring Ce moments is mediated by the spin 
polarization of conduction electrons at the Fermi level. When the conduction 
electron density is increased, the spin polarization would be enhanced, 
leading to the increase of $T_\mathrm{N}$. Another interesting observation in Fig. 
7(a) is the fact that both the suppression of $T_\mathrm{N}$ and increase of 
$\gamma $ value are more drastic as a function $y$ than as a function of $x$. When 
$5d$ holes are doped, the Fermi level is lowered toward the $4f$ electron level. 
Then, the \textit{c-f} exchange interaction will be strengthened and the valence 
fluctuation will be enhanced, as found in the case of Co substitution for Ni 
in CeNiSn.\cite{TT10, DT11} Another method to enhance the \textit{c-f} exchange interaction in Ce 
based compounds is the application of pressure. A previous study of 
CeOs$_{2}$Al$_{10}$ under pressure showed that $T_\mathrm{N}$ is almost constant up 
to $P = 2.3$ GPa and suddenly suppressed as the pressure is further 
increased.\cite{KU35} It is noteworthy that the suppression of $T_\mathrm{N}$ by pressure 
coincides with the metallization in $\rho (T)$ at $T < T_\mathrm{N}$, which 
resembles the observation in the present experiment by doping $5d$ holes. 

Finally, we focus on the relation between the hybridization gap and the AFM 
order. Figure 7(b) shows the variations of $T_\mathrm{N}$ and thermal activation 
energy $\mit\Delta$ in the resistivity as a function of $x$ and $y$. We find strong 
correlation between the variations of $T_\mathrm{N}$ and $\mit\Delta$. This correlation 
indicates that the presence of the hybridization gap is indispensable for 
the AFM order at unusually high $T_\mathrm{N}$. In order to prove this idea, 
we plan to do photoemission and electron-tunneling experiments 
which can probe the temperature dependence of the hybridization gap in Kondo 
semiconductors.\cite{TS36, TE37} 

\section{SUMMARY}
We studied the effects of doping of $5d$ electrons and holes on the Kondo 
semiconductor CeOs$_{2}$Al$_{10}$ by measuring $\chi (T)$, $M(B)$, $\rho (T)$, and 
$C(T)$ on single crystalline samples of 
Ce(Os$_{1-x}$Ir$_{x}$)$_{2}$Al$_{10}$($x \le  0.15$) and 
Ce(Os$_{1-y}$Re$_{y}$)$_{2}$Al$_{10}$($y \le  0.1$). The valence 
fluctuation behavior in $\chi_{a}(T)$ with a broad maximum at around 45 K 
for the undoped sample changes to the Curie-Weiss behavior of Ce$^{3+}$ with 
increasing $x$. This change means that doping of $5d$ electrons localizes the $4f$ 
electron state in CeOs$_{2}$Al$_{10}$. With increasing $y$, on the contrary, the 
broad maximum of $\chi_{a}(T)$ decreased and disappeared. The spin-flop 
transition in $M(B)$ for the Ir substituted samples revealed that the direction 
of the ordered moment $\mu_\mathrm{AF}$ changes from $\parallel c$-axis to $\parallel a$-axis with 
increasing $x$ to 0.03. In spite of the significant increase of $\mu_\mathrm{AF}$ 
from 0.3 $\mu_\mathrm{B}$/Ce for $x =0$ to 1 $\mu_\mathrm{B}/$Ce for $x =0.15$, 
$T_\mathrm{N}$ decreases gradually from 28.5 K to 7.0 K. By the Re substitution, 
$T_\mathrm{N}$ disappears at a small level $y = 0.05$. The results of $\rho (T)$ and 
$C(T)$ showed that the development of density of states at the Fermi level 
causes metallization at $x \cong 0.08$ and $y \cong 0.02$, respectively. 
Furthermore, the suppression of 
$T_\mathrm{N}$ is well correlated with that of gap energy $\mit\Delta$ as a function of $x$ and $y$. Therefore, 
we conclude that the presence of the hybridization gap is indispensable for 
the AFM order at unusually high $T_\mathrm{N}$ in CeOs$_{2}$Al$_{10}$.

\begin{acknowledgments}

We thank K. Yutani and Y. Yamada for their help in the preparation of 
samples. We acknowledge valuable discussions with T. Onimaru, D. T. Adroja, 
and T. Yokoya. This work was supported by a Grant-in-Aid of MEXT, Japan 
(Grant No. 20102004 and 23840033).

\end{acknowledgments}


\newpage

\begin{figure}
 \includegraphics[width=7cm]{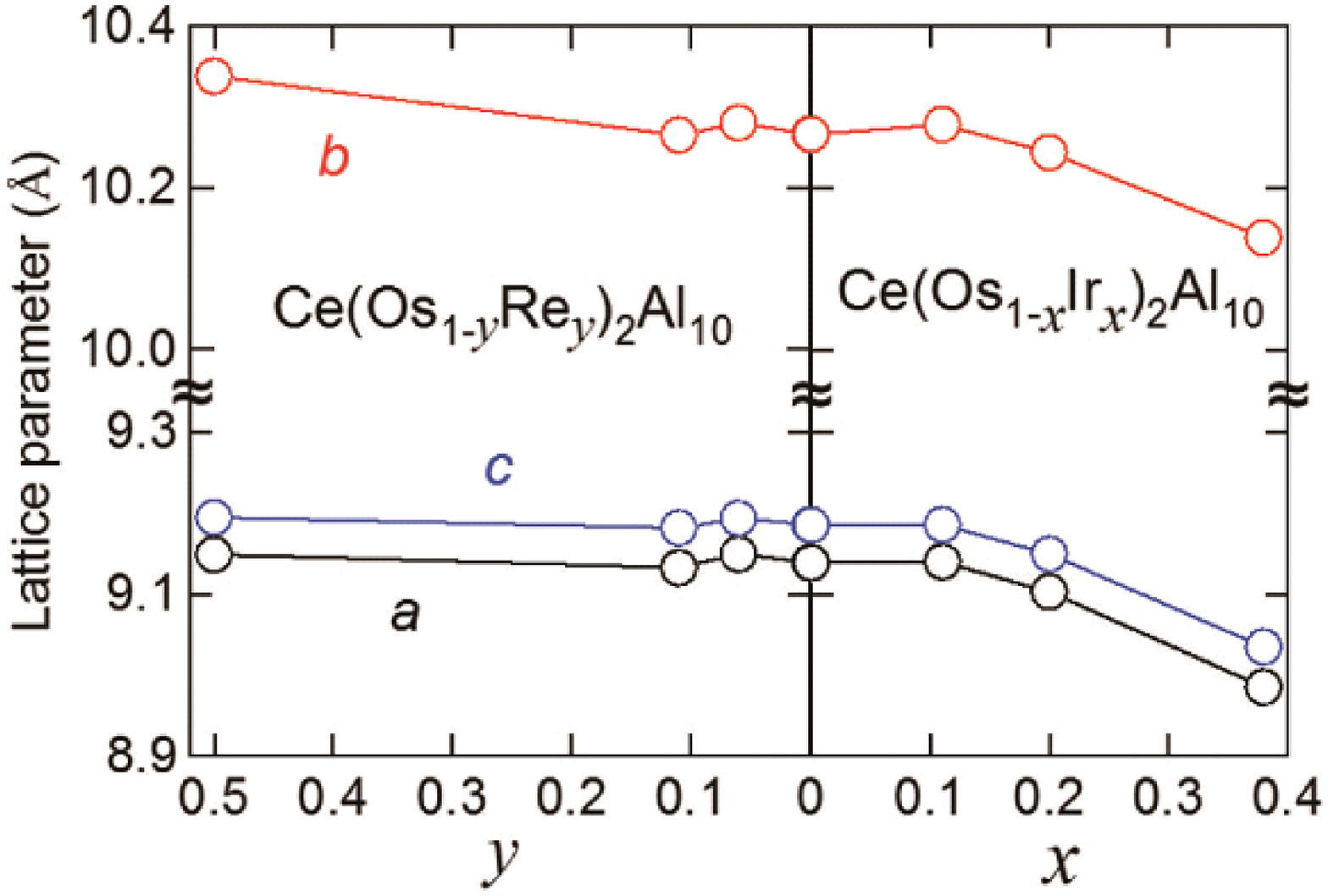}%
 \caption{\label{fig:epsart}Lattice parameters of polycrystalline samples of Ce(Os$_{1-x}$Ir$_{x}$)$_{2}$Al$_{10}$ and Ce(Os$_{1-y}$Re$_{y}$)$_{2}$Al$_{10}$, which crystalize in the orthorhombic YbFe$_{2}$Al$_{10}$-type structure.}
\end{figure}

\begin{figure}
 \includegraphics[width=7cm]{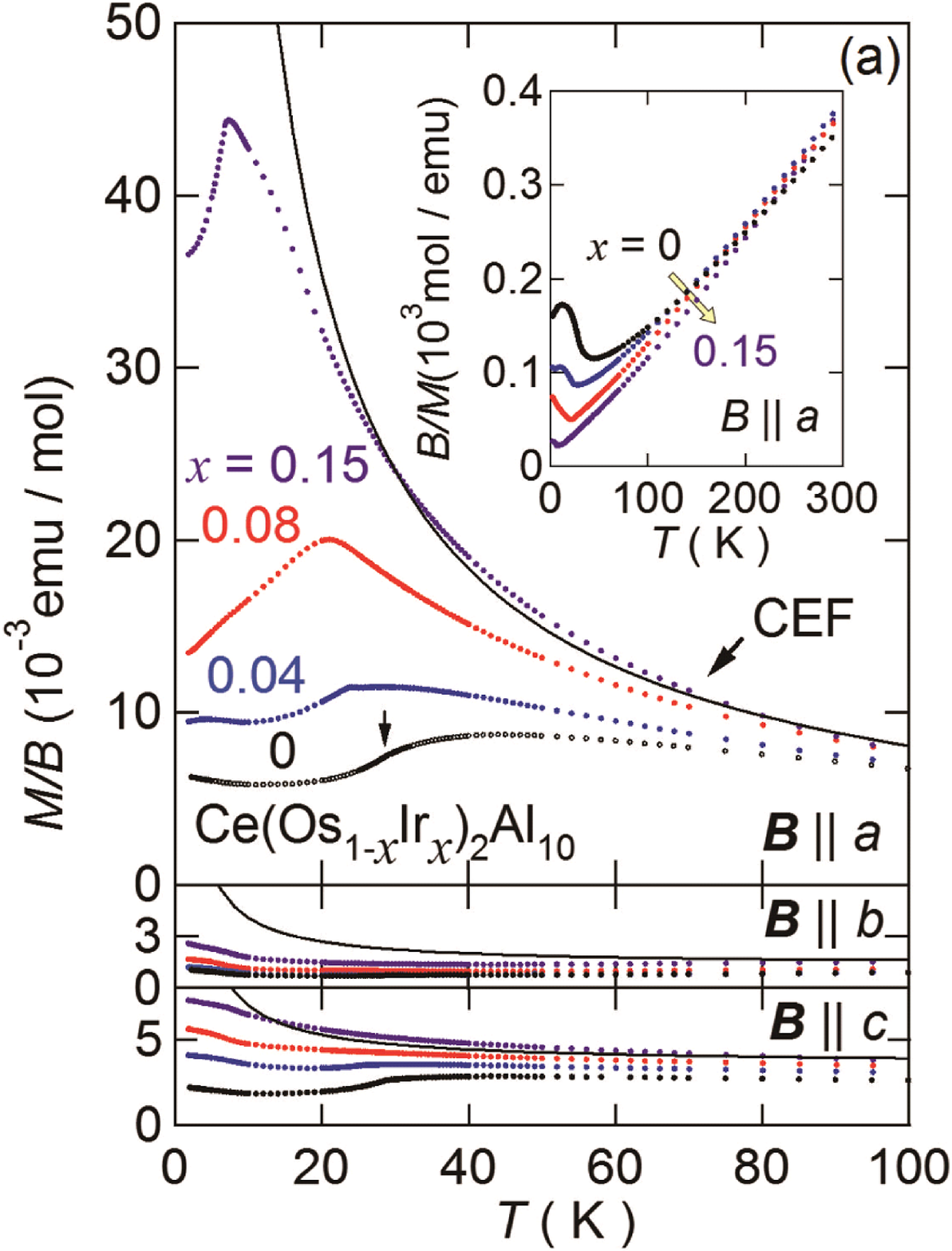}%
 \includegraphics[width=7cm]{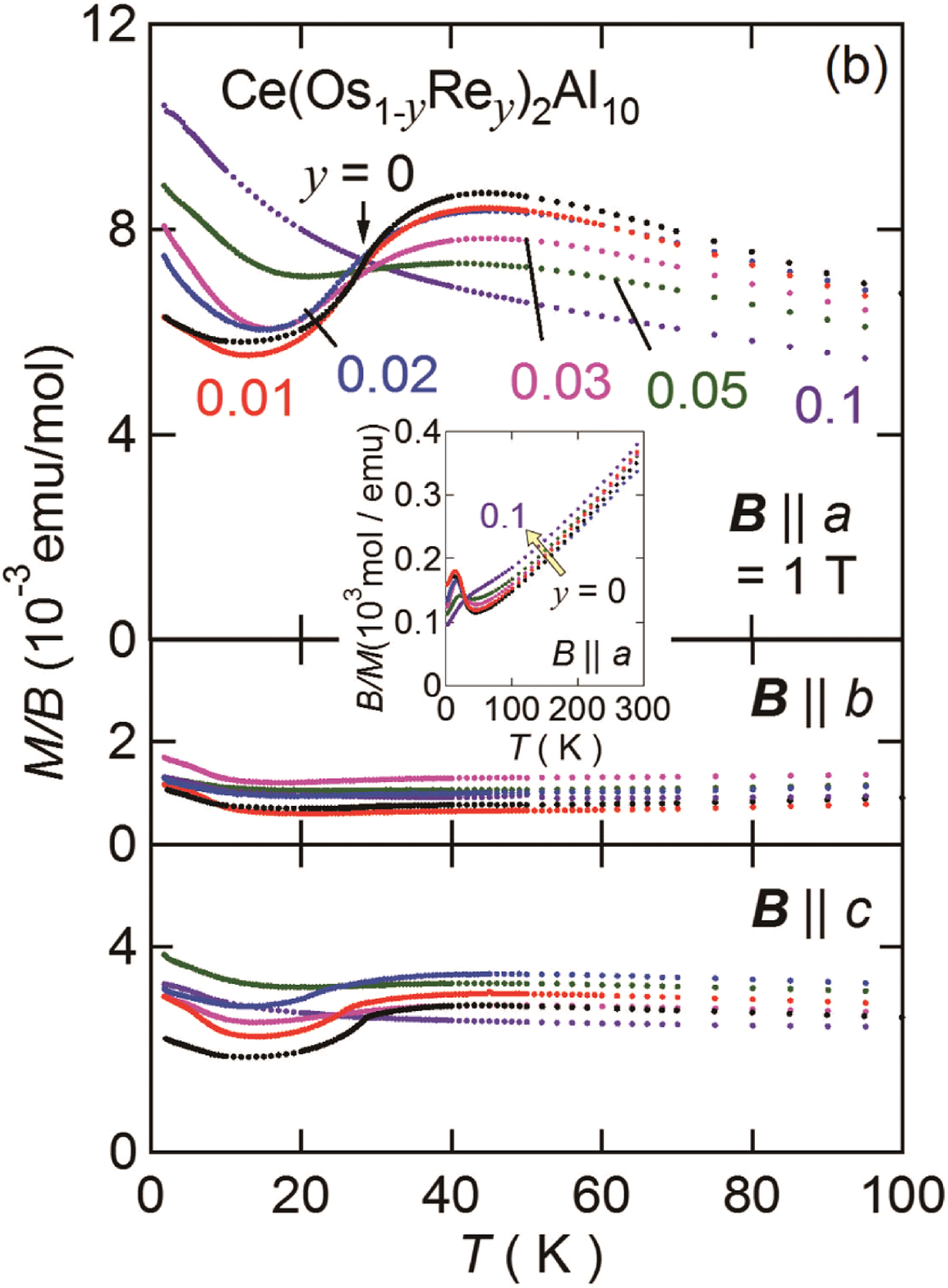}%
 \caption{\label{fig:epsart}Temperature dependence of magnetic susceptibility for single 
crystals of (a) Ce(Os$_{1-x}$Ir$_{x}$)$_{2}$Al$_{10}$ and (b) 
Ce(Os$_{1-y}$Re$_{y}$)$_{2}$Al$_{10}$ along the three principal axes.}
\end{figure}

\begin{figure}
 \includegraphics[width=7cm]{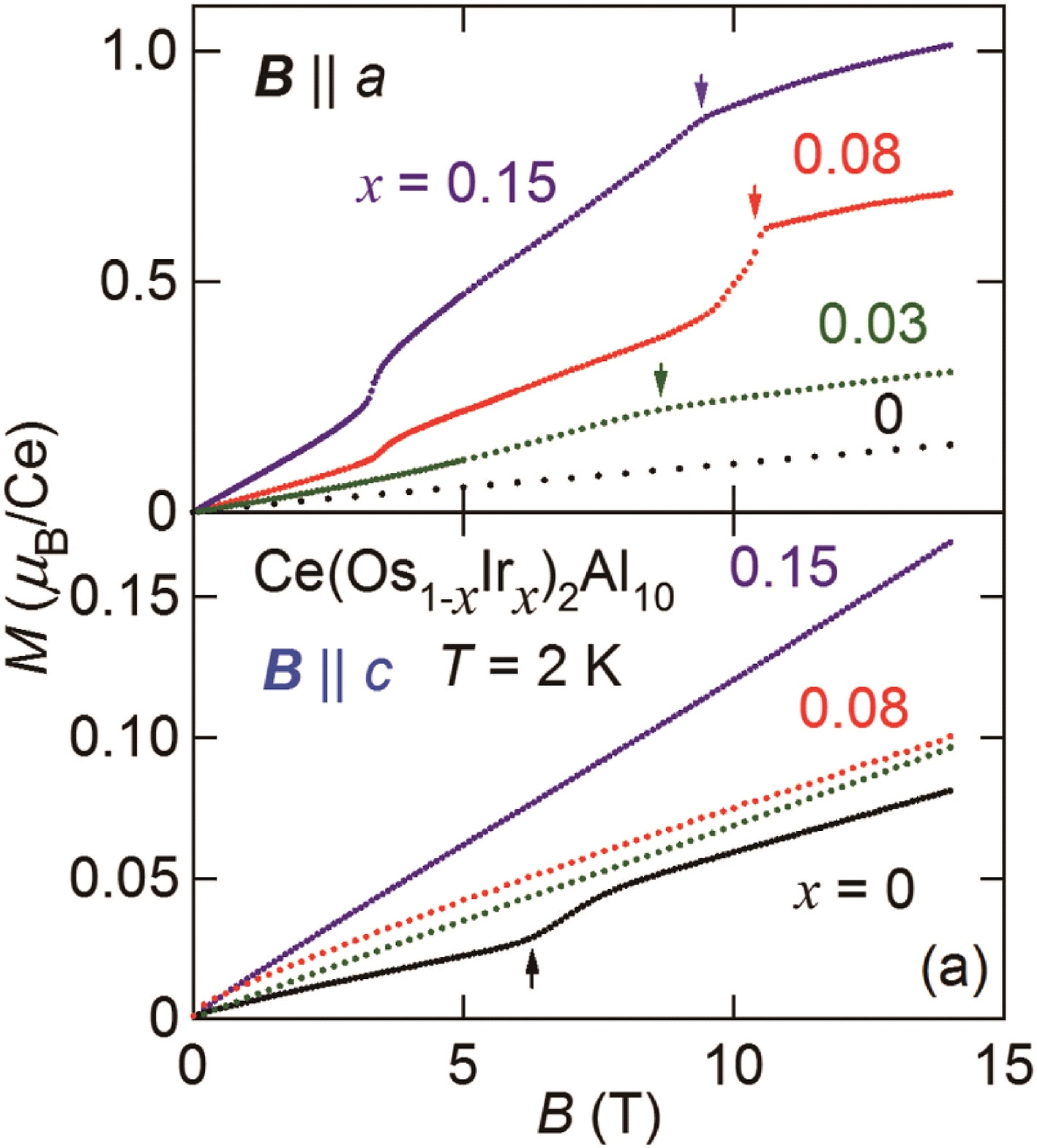}%
 \includegraphics[width=7cm]{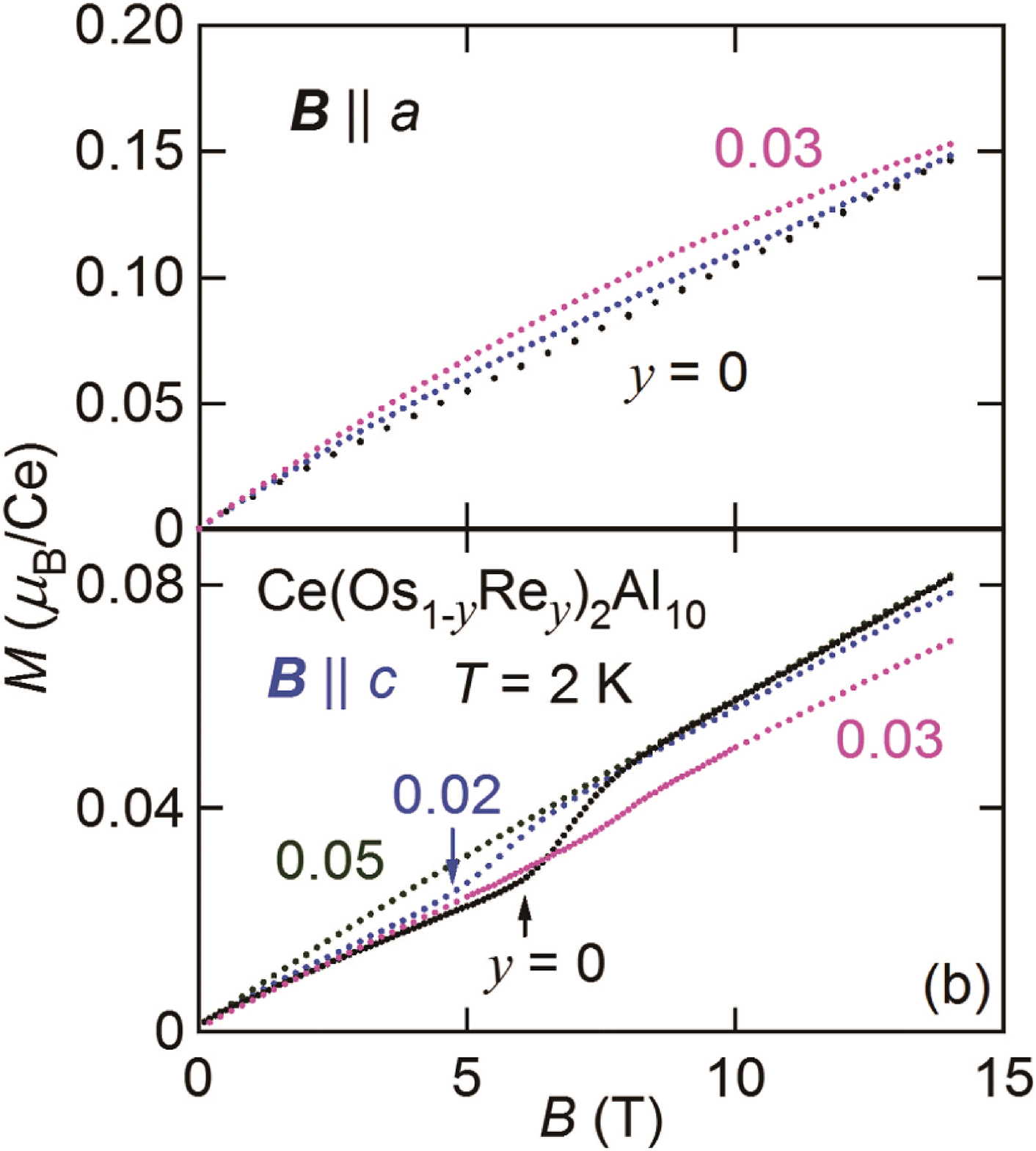}%
 \caption{\label{fig:epsart}Isothermal magnetization $M(B \parallel a)$ and $M(B \parallel c)$ at 2 K for (a) 
Ce(Os$_{1-x}$Ir$_{x}$)$_{2}$Al$_{10}$ and (b) 
Ce(Os$_{1-y}$Re$_{y}$)$_{2}$Al$_{10}$.}
\end{figure}

\begin{figure}
 \includegraphics[width=7cm]{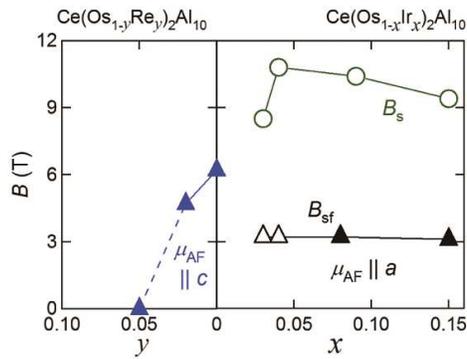}%
 \caption{\label{fig:epsart}Variations of the spin-flop field $B_{sf}$ and the field toward 
saturation $B_{s}$ in the isothermal magnetization curve $M(B)$ as a function of 
$x$ and $y$ in Ce(Os$_{1-x}$Ir$_{x}$)$_{2}$Al$_{10}$ and 
Ce(Os$_{1-y}$Re$_{y}$)$_{2}$Al$_{10}$.}
\end{figure}

\begin{figure}
 \includegraphics[width=7cm]{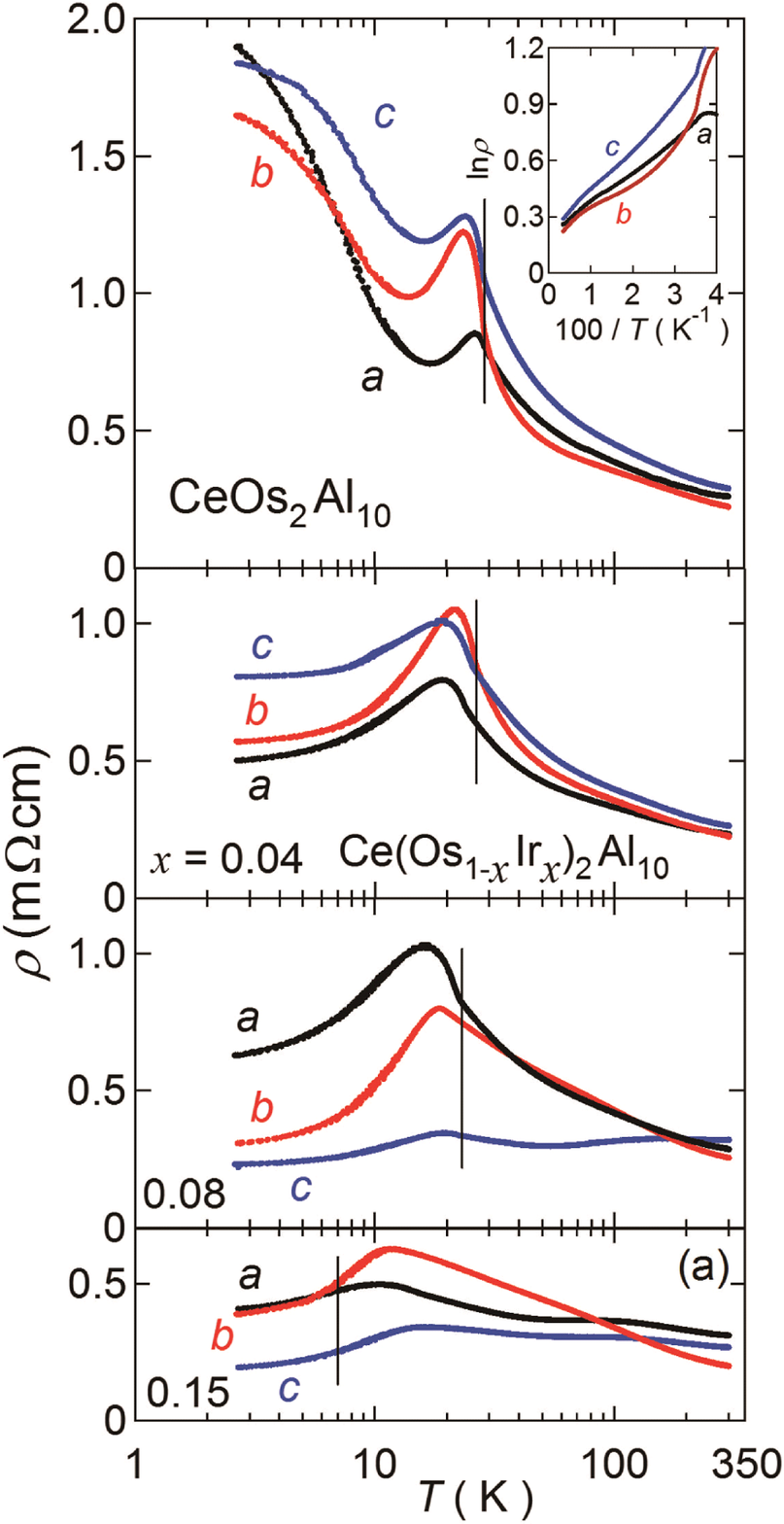}%
 \includegraphics[width=7cm]{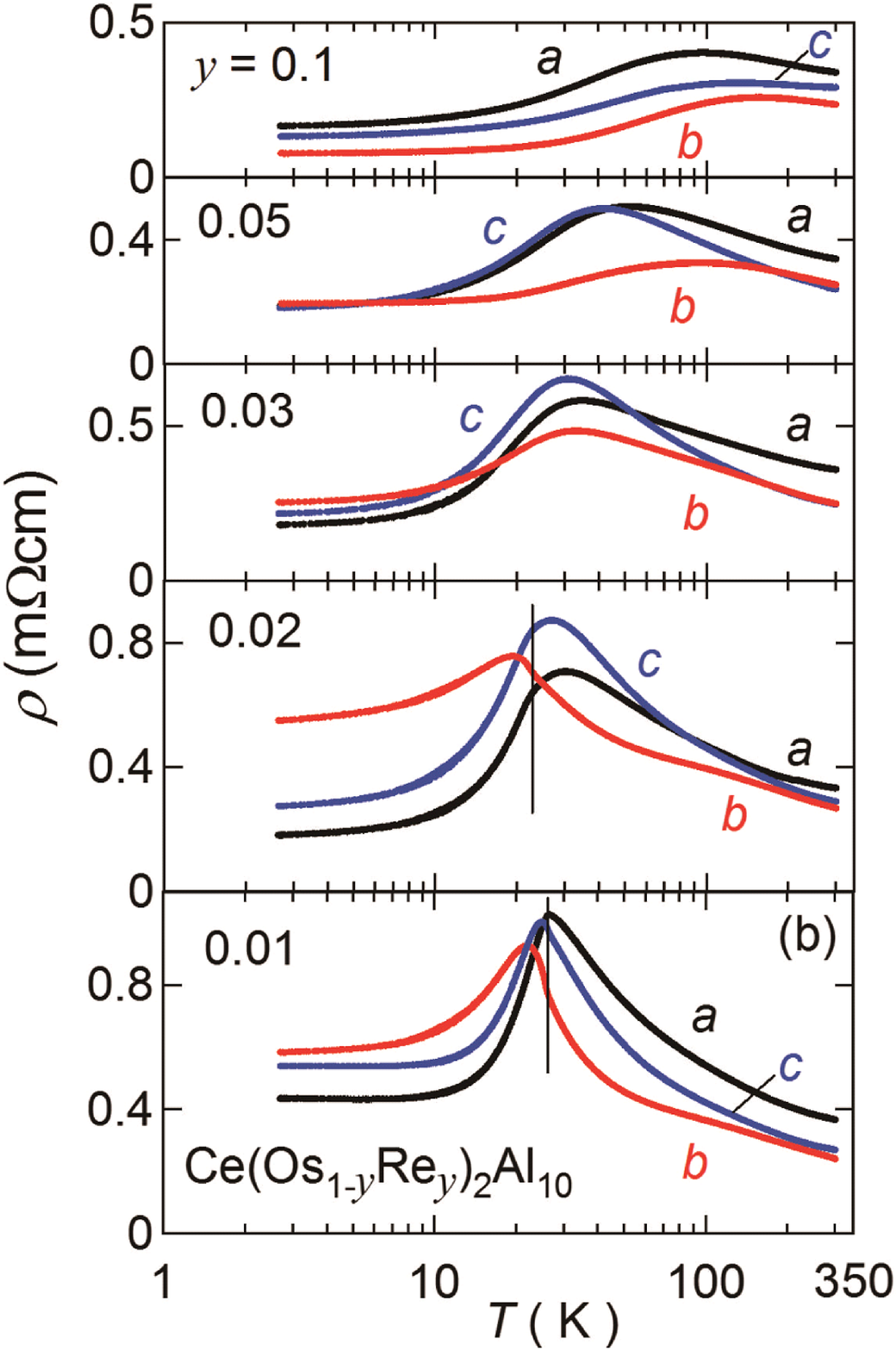}%
 \caption{\label{fig:epsart}Temperature dependence of electrical resistivity for single crystals 
of (a) Ce(Os$_{1-x}$Ir$_{x}$)$_{2}$Al$_{10}$ and (b) 
Ce(Os$_{1-y}$Re$_{y}$)$_{2}$Al$_{10}$ along the three principal axes.}
\end{figure}

\begin{figure}
 \includegraphics[width=7cm]{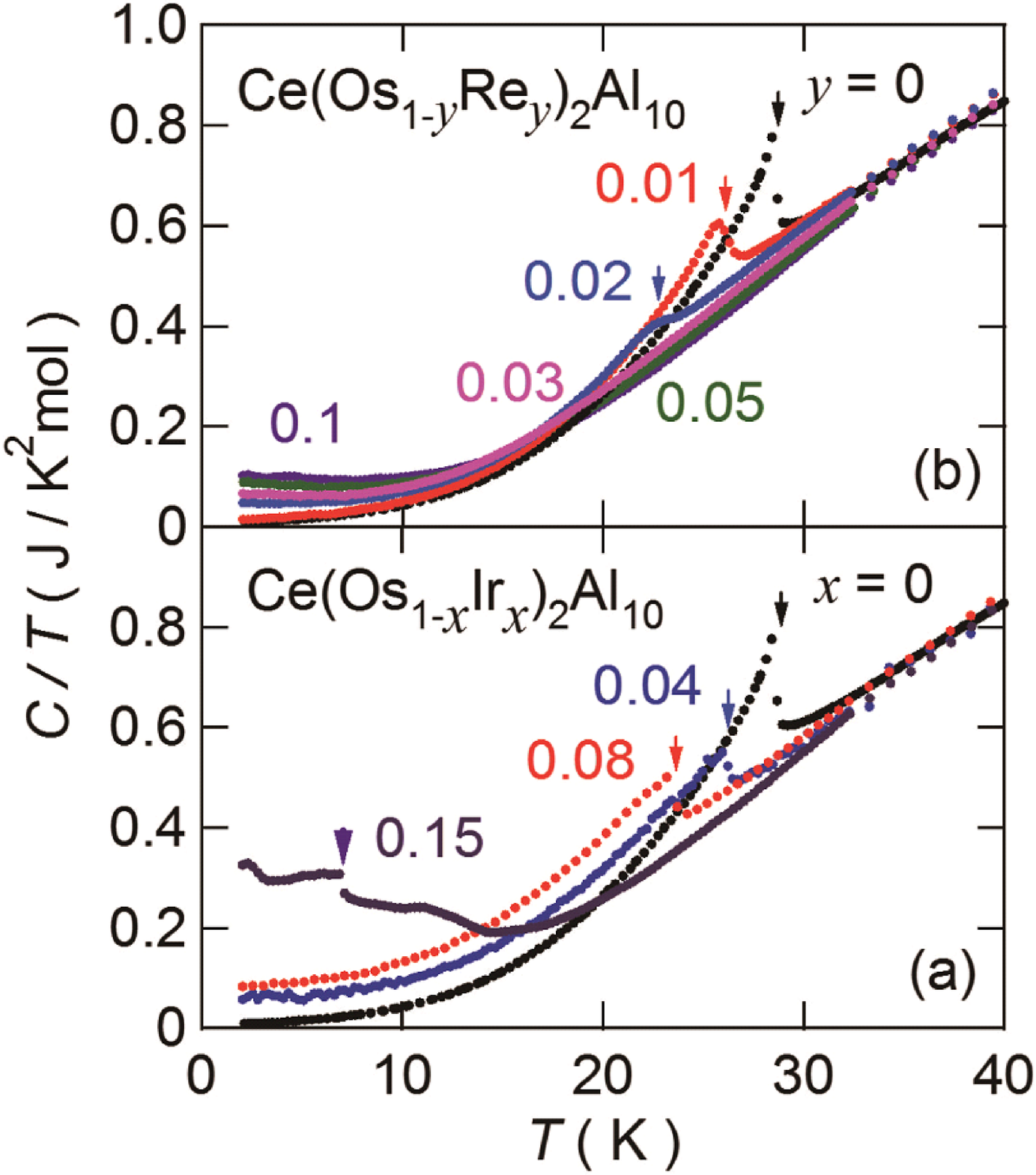}%
 \caption{\label{fig:epsart}Temperature dependence of specific heat divided by temperature 
$C/T$ for (a) Ce(Os$_{1-x}$Ir$_{x}$)$_{2}$Al$_{10}$ and (b) 
Ce(Os$_{1-y}$Re$_{y}$)$_{2}$Al$_{10}$.}
\end{figure}

\begin{figure}
 \includegraphics[width=7cm]{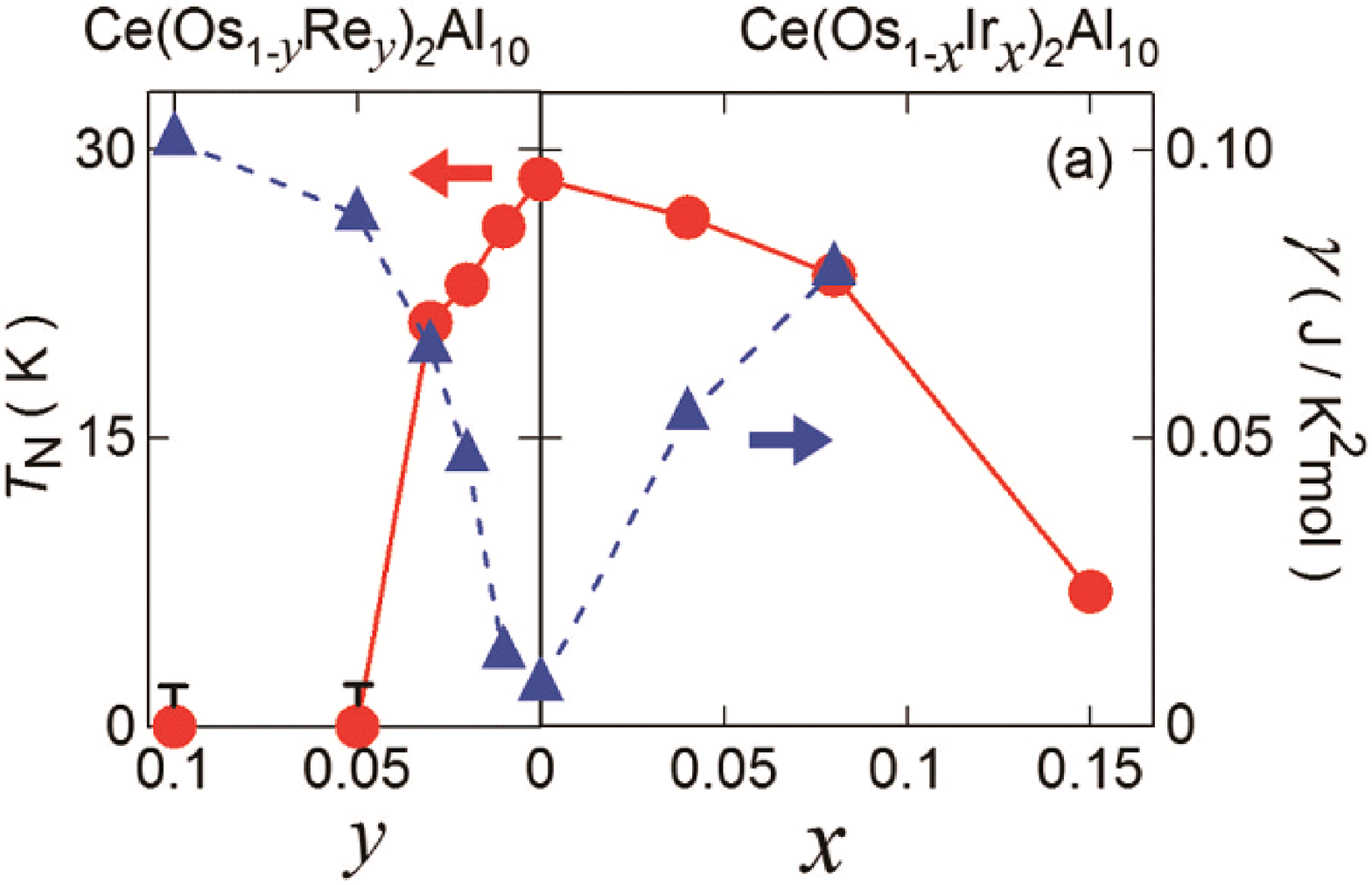}%
 \includegraphics[width=7cm]{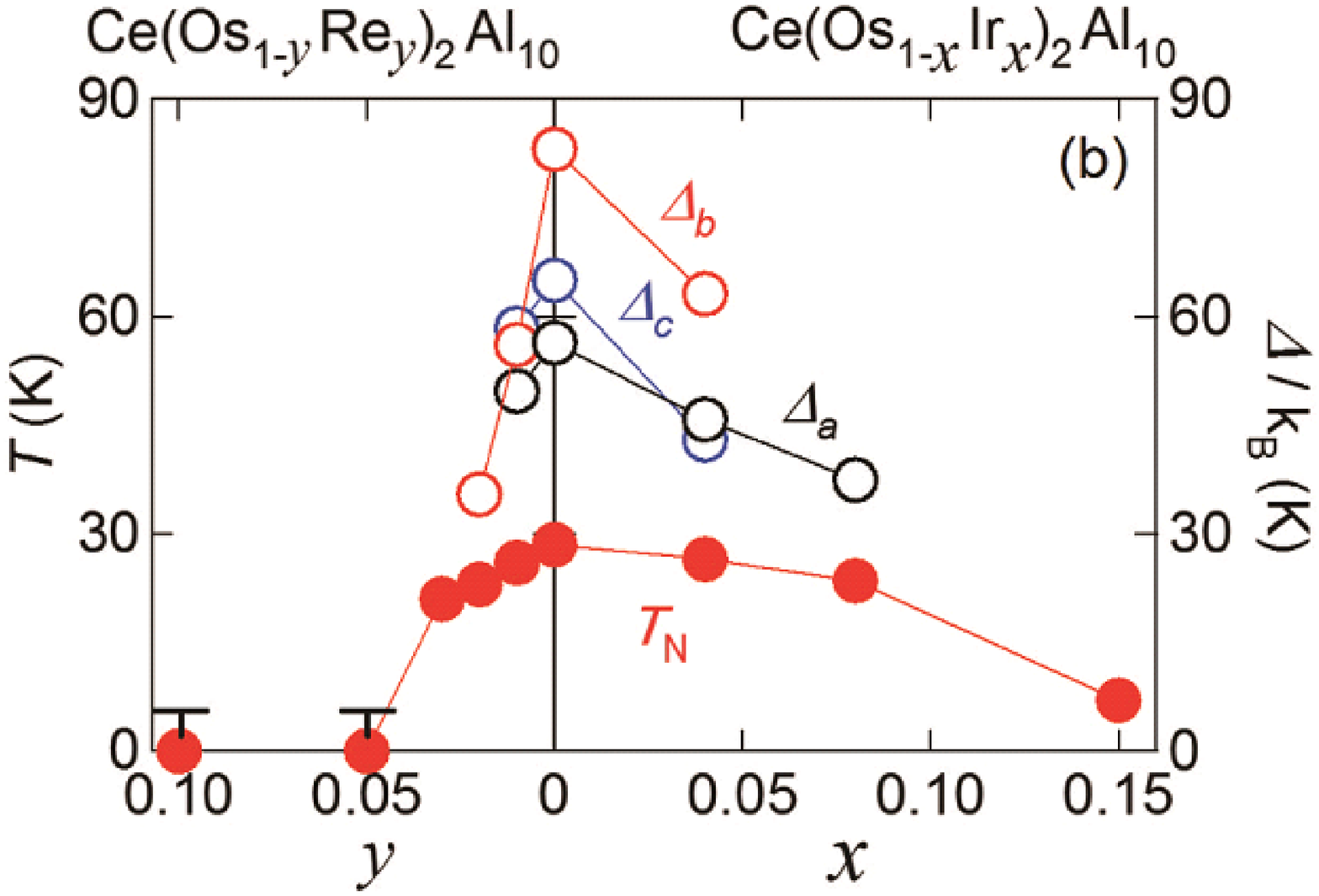}%
 \caption{\label{fig:epsart}Variations of (a) the N\'{e}el temperature $T_\mathrm{N}$ and 
the $\gamma$ value of the specific heat and (b) $T_\mathrm{N}$ and the thermal activation energy 
$\mit\Delta $ in the resistivity as a function of $x$ and $y$ in 
Ce(Os$_{1-x}$Ir$_{x}$)$_{2}$Al$_{10}$ and 
Ce(Os$_{1-y}$Re$_{y}$)$_{2}$Al$_{10}$.}
\end{figure}

\end{document}